\documentstyle[12pt]{article}

\begin{document}

\title{Dynamical disappearance of superposition states in the thermodynamic limit}

\author{Marco Frasca \\
        Via Erasmo Gattamelata, 3,\\
        00176 Roma (Italy)}

\date{\today}

\maketitle

\abstract{
It is shown that a macroscopic superposition state of radiation, strongly interacting
with an ensemble of two-level atoms, is removed generating a coherent state describing
a classical radiation field, when the thermodynamic limit is taken on the unitary
evolution obtained by the Schr\"odinger equation. Decoherence appears as a dynamical
effect in agreement with a recent proposal [M. Frasca, Phys. Lett. A {\bf 283}, 271 (2001)].
To prove that this effect is quite general, we show that this same behavior appears when
a superposition of two Fock number states is also considered.
Higher order corrections are computed showing that this result tends to
become exact in the thermodynamic limit. 
It appears as a genuine example of intrinsic collapse of the wave function.
}


\section{Introduction}

Recent experimental findings in mesoscopic devices \cite{exp} seem to suggest
that, at very low temperatures, some kind of mechanism is producing decoherence \cite{zur}.
These same experiments seems to prove that this mechanism is intrinsic to the
device. So, it is become demanding to understand a way to describe decoherence
at zero temperature \cite{moha1}.

Zero temperature decoherence in the framework of dissipative quantum systems has
been devised in \cite{loss1}. A general description of these kind of quantum
systems can be found in \cite{weiss}. In this paper we want to describe another
approach to decoherence, that is, decoherence produced by the unitary
evolution in the thermodynamic limit \cite{fra1}. This is non dissipative
decoherence, a first theory of which has been given in \cite{bon} where time
is considered as a stochastic variable. Besides, in mesoscopic physics,
some examples of non dissipative decoherence were also given in \cite{fra0,loss2}.

Some recent approaches rely on vacuum fluctuations to generate such kind of
decoherence \cite{butt} and, in mesoscopic devices it appears that electron-electron
interaction plays the dominant role. But, again these approaches can be leaded back
to dissipative quantum systems.

Non-dissipative decoherence can play a major role as it can make quantum theory
self-contained without requiring arbitrary separation between a bath and a
system. Here we demand that some systems, in the thermodynamics limit, develop
classical behavior. When such systems interact with other quantum systems, 
decoherence indeed appears. The evolution is anyhow unitary.

The foundation for this kind of view of decoherence is given by the formal analogy
between the time evolution operator of quantum mechanics written as
$\exp\left(-i\frac{Ht}{\hbar}\right)$ and the density matrix of a
system being at equilibrium $\exp\left(-\beta H\right)$. In the latter case the thermodynamic
limit has a precise meaning recovering standard thermodynamic \cite{kad}. Here
we will show that, in the case of quantum evolution, one can get classical
states when the system is properly prepared at the initial time \cite{fra1}.
Then, when such a system interacts with another quantum system, decoherence can
develop. Such classical states are characterized by having the system following
exactly the classical equations of motion, for given observables,
without any significant deviation
due to quantum fluctuations, in agreement with the Ehrenfest's theorem.

It is interesting to note that a recent experiment by Haroche and coworkers \cite{Har},
aimed at realizing a conceptual experiment on complementarity due to Bohr, proves
the appearance of classicality in the limit of increasing photons in a cavity. The
states they have produced are those theoretically predicted, in a pioneering
work on the Jaynes-Cummings model in the limit of a large number of photons, by
Gea-Banacloche \cite{gb}. In this latter work, the question of the collapse
of the wave function is properly discussed
in a similar context as ours, appearing as a first hint
toward the appearance of decoherence in the thermodynamic limit.

In this paper we want to limit our analysis to the particular case of a
single radiation mode interacting with $N$ two-level systems. So, by
studying the thermodynamic limit, we prove that, in the strong coupling
limit and with the two-level systems properly prepared, decoherence develops
destroying superposition states that are taken initially for the field mode,
driving the system toward a coherent state describing a classical field. 
Then, by computing higher order corrections, using perturbation theory,
we will be able to show how these terms are negligible small in the
thermodynamic limit, so that, the leading order term tends to become
an exact solution in the same limit. 
This seems a clear example of dynamical
collpse of the wave function obtained solving perturbatively the
Schr\"odinger equation.

The paper has the following structure. In sec \ref{sec1} we introduce the 
quantum model that we want to analyze. In sec.\ref{sec2} we show how the
appearance of classical states indeed happens for an ensemble of two-level systems. 
In sec.\ref{sec3} we give the leading order
solution of the model in the strong coupling regime proving that, given a
superposition of states, being Fock or coherent states, the system is driven
toward a coherent state describing a classical radiation field. In sec.\ref{sec5}
we evaluate the higher order corrections to the leading order solution to
show that, in the thermodynamic limit, our result does not change but rather
tends to be an exact solution. Finally, in
sec.\ref{sec6} we give the conclusions with a brief discussion.

\section{The model: two-level atoms interacting with a single radiation mode}
\label{sec1}

Our aim is to discuss how decoherence can appear as a dynamical effect, in the
limit of a very large number of systems (thermodynamic limit). It is a well-known
matter \cite{ct} that a single two-level atom under the effect of an increasing
number of radiation modes changes its behavior from coherent Rabi oscillations to
spontaneous emission, a typical decoherent effect.

Here, we reverse the situation by leaving a single radiation mode interacting 
with an ensemble of two-level atoms. The Hamiltonian of a single radiation mode
interacting with a two-level atom is given by \cite{ct}
\begin{equation}
    H = \omega a^\dagger a + \frac{\Delta}{2}\sigma_z + g\sigma_x(a+a^\dagger)
\end{equation}
being $\omega$ the frequency of the radiation mode,
$\Delta$ the separation between the levels of the atom, $g$ the coupling
between the radiation and the matter, $\sigma_x$ and $\sigma_z$ the Pauli matrices
and $a$ and $a^\dagger$ the annihilation and creation operators of the radiation mode.
%
In a recent experiment on a small Josephson-junction circuit 
irradiated with microwaves\cite{naka}, the strong
coupling regime for this Hamiltonian has been obtained describing Rabi
oscillations as theoretically described in \cite{fra2}. So, the strong
coupling regime has been practically realized in this physical situation.
%

This model
can be easily generalized to an ensemble of $N$ two-level atoms as
\begin{equation}
    H = \omega a^\dagger a + \sum_{i=1}^N \left[\frac{\Delta}{2}\sigma_{zi} 
	    + g\sigma_{xi}(a+a^\dagger)\right] \label{eq:myH}
\end{equation}
and again we consider the strong coupling regime as, in this case, decoherence
is produced dynamically by unitary evolution in the thermodynamic limit. 
This
model has been recently studied, in a similar context, in Ref.\cite{sola}.
%
It can be realized by an array of Josephson-junction circuits, irradiated by
microwaves, working as qubits for quantum computer.

Let us note, at this point, that having our Hamiltonian the form
\begin{equation}
    H = H_{system} + H_{bath} + H_{system-bath},
\end{equation}
we are neglecting $H_{bath}$, that is, a dual situation with respect to the
one considered by Haake and coworkers \cite{haa} that neglect $H_{system}$. But,
while they describe dissipative decoherence, we aim to show, using perturbation
theory, that decoherence is produced dynamically in the thermodynamic limit.

\section{Classical states by unitary evolution and thermodynamic limit}
\label{sec2}

In this section we want to limit our study to the Hamiltonian of the ensemble of
two-level atoms
\begin{equation}
    H_0 = \sum_{i=1}^N \frac{\Delta}{2}\sigma_{zi}
\end{equation}
and we prove the following result \cite{fra1}:

{\sl For a proper set of initial conditions, the ensemble of two-level atoms
evolves in time classically with respect to the variables 
$\Sigma_x=\sum_{i=1}^N\sigma_{xi}$,
$\Sigma_y=\sum_{i=1}^N\sigma_{yi}$,
$\Sigma_z=\sum_{i=1}^N\sigma_{zi}$, in the thermodynamic limit.}

We want to emphasize that the system does not decohere but evolves in time
washing out the quantum fluctuations in the thermodynamic limit. Besides, the
system must be properly prepared as we cannot claim that when the system is in
an eigenstate of $H_0$, behaves classically. 

To prove our result, we consider the following initial state
\begin{equation}
    |\psi(0)\rangle = \prod_{i=1}^N\left(a_i|\downarrow\rangle_i + b_i|\uparrow\rangle_i\right) \label{eq:is}
\end{equation} 
with $|a_i|^2+|b_i|^2=1$. The set of coefficients $a_i$ and $b_i$ must be chosen in
such a way that $|\psi(0)\rangle$ is not in an eigenstate of $H_0$. The time
evolution is determined through
\begin{equation} 
U_0(t)=\prod_{i=1}^N[e^{it\frac{\Delta }{2}}|\downarrow\rangle_i \ _i \langle\downarrow|+
e^{-it\frac{\Delta }{2}}|\uparrow\rangle_i \ _i \langle\uparrow|] \label{eq:u}
\end{equation}
giving
\begin{equation} 
|\psi(t)\rangle=
\prod_{i=1}^N(a_i e^{it\frac{\Delta }{2}}|\downarrow\rangle_i+
b_ie^{-it\frac{\Delta }{2}}|\uparrow\rangle_i).
\end{equation} 
The average values are given by
\begin{equation} 
\langle \Sigma_x \rangle=\langle\psi(t)| \Sigma_x |\psi(t)\rangle
=N(\xi_H \cos(\Delta t)+\xi'_H \sin(\Delta t))
\end{equation} 
being
$\xi_H=\sum_{i=1}^N(a^*_i b_i + a_i b^*_i)/N$ and
$\xi'_H=i \sum_{i=1}^N(a^*_i b_i - a_i b^*_i)/N$, 
numbers of order of unity. In the same way one has
\begin{equation} 
\langle \Sigma_x^2 \rangle=\langle\psi(t)| \Sigma_x^2 |\psi(t)\rangle=
N\left[1-\frac{1}{N}\sum_{i=1}^N(a_i^*b_ie^{i\Delta t}+b_i^*a_ie^{-i\Delta t})^2\right]+
N^2(\xi_H \cos(\Delta t)+\xi'_H \sin(\Delta t))^2
\end{equation}
so that, one gets finally that $(\Delta\Sigma_x)^2\propto N$ and the mean values are
overwhelming large with respect to quantum fluctuations in the thermodynamic
limit. A similar argument runs for $\Sigma_y$ and $\Sigma_z$ components.
This means in turn that
the classical equations of motion from the Ehrenfest theorem
\begin{eqnarray}
    \langle \dot\Sigma_x \rangle &=& -\Delta \langle \Sigma_y \rangle \\ \nonumber
	\langle \dot\Sigma_y \rangle &=& \Delta \langle \Sigma_x \rangle \\ \nonumber
	\langle \dot\Sigma_z \rangle &=& 0
\end{eqnarray}
are obeyed without any significant deviation due to quantum fluctuations
in the limit of very large $N$. This proves our assertion. 

A relevant question that can be asked with this result is if such a model,
with this kind of initial state, can produce decoherence interacting with
other quantum systems. We will answer this question in the next sections.

\section{Decoherence generated by unitary evolution: Leading order solution}
\label{sec3}

We want to show as, an ensemble of two-level atoms, strongly interacting with
a single radiation mode, produces decoherence dynamically in the thermodynamic
limit. For our aim, we start initially with a macroscopic quantum superposition
state (generally known in literature as a Scrh\"odinger cat states)
given by \cite{sch}
\begin{equation}
    |\psi(0)\rangle = {\cal N}(|\alpha e^{i\phi}\rangle + |\alpha e^{-i\phi}\rangle)|\chi\rangle
\end{equation}
being $|\chi\rangle$ the contribution of the two-level atoms to be specified further,
$\alpha$ a real number, $\phi$ a phase and $\cal N$ a normalization factor
given by
\begin{equation}
    {\cal N}^2 = \frac{1}{2+2\cos[\alpha^2\sin(2\phi)]\exp(-2\alpha^2\sin^2\phi)},
\end{equation}
accounting for the nonorthogonality of the two coherent states. 

Then, we consider the strong coupling regime of the Hamiltonian (\ref{eq:myH})
as done in Ref.\cite{fra2} by assuming as an unperturbed Hamiltonian
\begin{equation}
     H_u = \omega a^\dagger a + g\sum_{i=1}^N\sigma_{xi}(a+a^\dagger).
\end{equation}
We can compute the unitary evolution operator by treating formally $\Sigma_x = \sum_{i=1}^N\sigma_{xi}$
as a c-number obtaining
\begin{equation}
    U_F(t) =  e^{i\hat\xi(t)}e^{-i\omega a^\dagger at}
	\exp[\hat\beta(t)a^\dagger - \hat\beta(t)^\dagger a],
\end{equation}
being
\begin{equation}
    \hat\xi(t)=\frac{\Sigma_x^2g^2}{\omega^2}(\omega t - \sin(\omega t))
\end{equation}
and
\begin{equation}
    \hat\beta(t)=\frac{\Sigma_x g}{\omega}(1-e^{i\omega t})
\end{equation}
operators for the ensemble of two-level atoms.

At this stage, we choose a simplified form of the initial state (\ref{eq:is}) by simply taking
the eigenstate $|1\rangle = \frac{1}{\sqrt{2}}(|\downarrow\rangle + |\uparrow\rangle)$ of $\sigma_x$,
with eigenvalue $1$, for each atom in the ensemble, so
\begin{equation}
    |\chi\rangle = \prod_{i=1}^N|1\rangle_i.
\end{equation}
This kind of ``ferromagnetic'' state simplifies the computation giving us at the leading order in the
strong coupling perturbation series \cite{fra3}
\begin{equation}
     |\psi(t)\rangle\approx U_F(t)|\psi(0)\rangle = e^{i\xi(t)}
	 {\cal N}(e^{i\phi_1(t)}|\beta(t)e^{-i\omega t} 
	 + \alpha e^{i\phi-i\omega t}\rangle 
	 + e^{i\phi_2(t)}|\beta(t)e^{-i\omega t} + \alpha e^{-i\phi-i\omega t}\rangle)|\chi\rangle. \label{eq:st}
\end{equation}
We have used the property of the displacement operator for coherent states so to yield
\begin{equation}
    \xi(t)=\frac{N^2g^2}{\omega^2}(\omega t - \sin(\omega t)),
\end{equation}
\begin{equation}
    \beta(t) = \frac{Ng}{\omega}(1-e^{i\omega t})
\end{equation}
and
\begin{equation}
    \phi_1(t) = -i\frac{\alpha}{2}[\beta(t)e^{-i\phi}-\beta^*(t)e^{i\phi}],
\end{equation}
\begin{equation}
    \phi_2(t) = -i\frac{\alpha}{2}[\beta(t)e^{i\phi}-\beta^*(t)e^{-i\phi}],
\end{equation}
with the phases $\phi_1(t)$ and $\phi_2(t)$ generated by the multiplication of two displacement operators.
Then, it is straightforward to impose the thermodynamic limit $N\rightarrow\infty$, 
keeping $\alpha$ fixed, to verify that the
macroscopic quantum superposition state is driven to the classical state $|\beta(t)e^{-i\omega t}\rangle$,
proving our assertion. The system decoheres removing the superposition of the states. This state describes
a classical radiation field as we are in the thermodynamic limit and so, the quantum fluctuations can be
neglected with respect to the mean value \cite{mw}.

One may ask if this is true decoherence. Actually, looking at the state (\ref{eq:st}) it
appears as if we have just displaced the initial coherent state and it seems that the
quantum effects are all there yet and we have not disposed of them as normally happens
by a true decoherent model. We prove now, by computing the Wigner function of the state
(\ref{eq:st}), that, in the thermodynamic limit, quantum effects are washed away and
decoherence is recovered. Indeed, it is straightforward to obtain
just for
the interference
term containing negativity and oscillations as
\begin{eqnarray}
    W_{INT} &=&\frac{2}{\pi}
	\exp\left[-\left(x + \frac{\sqrt{2}Ng}{\omega}(1-\cos(\omega t)) - \sqrt{2}\alpha\cos(\phi)\cos(\omega t)\right)^2\right] \\ \nonumber
	&\times&\exp\left[-\left(p + \frac{\sqrt{2}Ng}{\omega}\sin(\omega t) + \sqrt{2}\alpha\cos(\phi)\sin(\omega t)\right)^2\right] \\ \nonumber
	&\times&\cos\left[2\sqrt{2}\alpha\sin(\phi)\left(p\sin(\omega t) - x\cos(\omega t)\right)
	+\alpha^2\sin(2\phi) + 
	8\alpha\frac{Ng}{\omega}\sin(\phi)(1-\cos(\omega t))\right].
\end{eqnarray}
It is easily realized that it depends also on time that appears in strongly oscillating
terms like $\cos\left(8\alpha\frac{Ng}{\omega}(1-\cos(\omega t))\sin(\phi)\right)$
and $\sin\left(8\alpha\frac{Ng}{\omega}(1-\cos(\omega t))\sin(\phi)\right)$ and a
sense should be attached to such terms in the thermodynamic limit $N\rightarrow\infty$.
Indeed, this is mathematically possible and such terms can be assumed to be zero in
such a limit, e.g. in the sense of Abel or Euler \cite{hardy}.
%
%
A physical meaning should be attached to the oscillating part taken to be zero in the
thermodynamic limit. Indeed,
%
%
One can see that the time scale of variation of the oscillating part becomes even more
smaller as $N$ increases. This means that one could reach, in principle, oscillations on
a time scale of the Planck time but one should expect that an average in time happens
largely before this can happen. This is the question of the singular limits that are
at the foundations of the theory of decoherence as pointed out by Berry\cite{berry}.  
So, the interference term can be neglected and true decoherence happens.
This point of view has also been emphasized in Ref. \cite{fra0} for a spin 
interacting with a spin bath. So, we recognize commonality between these models
and
the limit in the Abel or Euler sense translates into an average in time.
%

We want to verify if this same effect happens for a superposition of number Fock states. So, we
consider an initial state given by
\begin{equation}
    |\psi(0)\rangle = \frac{1}{\sqrt{2}}(|0\rangle + |k\rangle)|\chi\rangle
\end{equation}
being $k$ an integer, yielding
\begin{equation}
    |\psi(t)\rangle \approx U_F(t)|\psi(0)\rangle = e^{i\xi(t)}
	\frac{1}{\sqrt{2}}\left(|\beta(t)e^{-i\omega t}\rangle + e^{-ik\omega t}
	|k,\beta(t)e^{-i\omega t}\rangle\right)|\chi\rangle \label{eq:sol}
\end{equation}
being
\begin{equation}
    |k,\beta(t)e^{-i\omega t}\rangle = \sum_{n=0}^\infty |n\rangle\langle n|D[\beta(t)e^{-i\omega t}]|k\rangle
\end{equation}
a displaced number state \cite{kn}. From Ref.\cite{kn} we derive the following formulas
\begin{equation}
    \langle n|D[\alpha]|k\rangle = \left(\frac{k!}{n!}\right)^\frac{1}{2}\alpha^{(n-k)}
	e^{-\frac{|\alpha|^2}{2}}L_k^{(n-k)}(|\alpha|^2)
\end{equation}
for  $n\ge k$, otherwise one has
\begin{equation}
    \langle n|D[\alpha]|k\rangle = \left(\frac{n!}{k!}\right)^\frac{1}{2}(-\alpha^*)^{(k-n)}
	e^{-\frac{|\alpha|^2}{2}}L_n^{(k-n)}(|\alpha|^2)
\end{equation}
being $L_n^k(x)$ the associated Laguerre polynomials \cite{gr}. We put
\begin{equation}
    \beta'(t) = \beta(t)e^{-i\omega t}
\end{equation}
and then rewrite eq.(\ref{eq:sol}) as
\begin{eqnarray}
    |\psi(t)\rangle &\approx& e^{i\xi(t)}
	\frac{1}{\sqrt{2}}\left(|\beta'(t)\rangle + \right.\\ \nonumber
	& &e^{-ik\omega t}\sum_{n=0}^{k-1}|n\rangle \left(\frac{n!}{k!}\right)^\frac{1}{2}[-\beta'^*(t)]^{(k-n)}
	e^{-\frac{|\beta'(t)|^2}{2}}L_n^{(k-n)}(|\beta'(t)|^2) + \\ \nonumber
	& &\left.e^{-ik\omega t}\sum_{n=k}^\infty |n\rangle \left(\frac{k!}{n!}\right)^\frac{1}{2}[\beta'(t)]^{(n-k)}
	e^{-\frac{|\beta'(t)|^2}{2}}L_k^{(n-k)}(|\beta'(t)|^2)
	\right)|\chi\rangle. \label{eq:sol2}
\end{eqnarray}
In the thermodynamic limit we can take $\beta'(t)$ to become increasingly large
being proportional to $N$. So, we can approximate the associated Laguerre polynomials
as $L_n^k(x)\approx \frac{(-x)^n}{n!}$ for $x\rightarrow\infty$ and substituting
into eq.(\ref{eq:sol2}) one gets
\begin{equation}
    |\psi(t)\rangle \approx e^{i\xi(t)}
	\frac{1}{\sqrt{2}}\left(|\beta'(t)\rangle + e^{-ik\omega t}\frac{[-\beta'^*(t)]^k}{\sqrt{k!}}
	\sum_{n=0}^\infty |n\rangle \frac{1}{\sqrt{n!}}[\beta'(t)]^n
	e^{-\frac{|\beta'(t)|^2}{2}}
	\right)|\chi\rangle
\end{equation}
and it is not difficult to recognize the second term on the r.h.s. being again the
coherent state $|\beta'(t)\rangle$. Then, the radiation field is driven again
toward a coherent state describing a classical field, the initial superposition
state being washed out in the thermodynamic limit.

In both cases we are left with corrections to the normalization factor due to the
approximations we have introduced. They will not play any role when the average
values are computed
as we have applied a limit procedure on the wave function implying a
recomputation of the normalization factor.

\section{Decoherence generated by unitary evolution: Higher order corrections}
\label{sec5}

In this section our aim will be to compute higher order corrections to the
results in sec.\ref{sec3} to prove that indeed, these corrections do not
modifiy our argument in the given approximations.

The dual Dyson series can be formally written as \cite{fra3}
\begin{equation}
    U(t) = U_F(t)T\exp\left(-i\int_{t_0}^tdt' U_F^\dagger(t')H_0U_F(t')\right)
\end{equation}
being $T$ the time ordering operator, giving at the first order
\begin{equation}
    U^{(1)}(t) = -iU_F(t)\int_{t_0}^tdt' U_F^\dagger(t')H_0U_F(t') \label{eq:fo}
\end{equation}
and, for the second order
\begin{equation}
    U^{(2)}(t) = -U_F(t)\int_{t_0}^tdt' U_F^\dagger(t')H_0U_F(t')\int_{t_0}^{t'}dt'' U_F^\dagger(t'')H_0U_F(t'').
\end{equation}
Our aim is to evaluate both the contributions. The first order solution has
been put forward, firstly, in Ref.\cite{fra2}. We introduce a quite general way
to evaluate higher order corrections by introducing the reduced unitary
evolution as
\begin{equation}
    U_{R\chi}(t)=\langle\chi|U(t)|\chi\rangle.
\end{equation}
We consider only the transitions with the spin bath untouched, the most relevant
for our analysis. We will justify {\sl a posteriori} this definition. This
describes an unitary evolution for the radiation field. As we have seen in
Sec.\ref{sec3}, the unitary evolution, at the leading order,
produces the disappearance of superposition states in the thermodynamic limit. 
This effect, also true for the reduced unitary evolution,
disappears for all the other orthogonal states of the spin bath
that appear during the evolution.

Indeed, by eq.(\ref{eq:fo}) the first order gives
\begin{equation}
    -ie^{i\frac{(N-2)^2g^2}{\omega^2}(\omega t -\sin(\omega t))}
	e^{-i\omega a^\dagger at}e^{\frac{N-2}{N}[\beta(t)a^\dagger-\beta(t)^*a]}
	\frac{\Delta}{2}
	\int_0^tdt'e^{4i(N-1)\frac{g^2}{\omega^2}(\omega t'-\sin(\omega t'))}
	e^{\alpha(t')a^\dagger-\alpha(t')^*a}|\hat\chi'\rangle
\end{equation}
with
\begin{equation}
    \alpha(t) = \frac{2\beta(t)}{N},
\end{equation}
not dependent on $N$.
But the first order correction modifies the spin bath state as
\begin{equation}
    |\hat\chi'\rangle = |-1\rangle_1|1\rangle_2\cdots|1\rangle_N+
	|1\rangle_1|-1\rangle_2\cdots|1\rangle_N+\cdots+
	|1\rangle_1|1\rangle_2\cdots|-1\rangle_N,
\end{equation}
that is a non normalized state orthogonal to $|\chi\rangle$. We normalize this
state by multiplying it by the factor $\frac{1}{\sqrt{N}}$ and call it $|\chi'\rangle$.
The conclusion is
that the first order term is zero and does not contribute to the reduced unitary evolution. But, if we
want to know the reduced unitary evolution with respect to the state $|\chi'\rangle$ we
can see that now is the leading order that does not contribute being zero, and we have
\begin{eqnarray}
    U^{(1)}_{R\chi'}(t) &=& \langle\chi'|U^{(1)}(t)|\chi\rangle = \\ \nonumber
	& &-ie^{i\frac{(N-2)^2g^2}{\omega^2}(\omega t -\sin(\omega t))}
	e^{-i\omega a^\dagger at}e^{\frac{N-2}{N}[\beta(t)a^\dagger-\beta(t)^*a]}\sqrt{N}
	\frac{\Delta}{2}\times \\ \nonumber
	& &\int_0^tdt'e^{4i(N-1)\frac{g^2}{\omega^2}(\omega t'-\sin(\omega t'))}
	e^{\alpha(t')a^\dagger-\alpha(t')^*a}.
\end{eqnarray}
This term, in the thermodynamic limit $N\rightarrow\infty$, has the integral wildly 
oscillating and giving a contribution proportional to $\frac{1}{N}$. So, this term
that modifies the unitary evolution as due to a modified state of the spin bath, can
be safely neglected in the thermodynamic limit, being $O\left(\frac{1}{\sqrt{N}}\right)$. 
This proves {\sl a posteriori}, as promised, our statement on the relevance 
of the reduced unitary evolution just
with respect to the state $|\chi\rangle$. Other non orthogonal states appear
at higher orders, as we will see, but their contribution is even more smaller in
the thermodynamic limit.

We can now evaluate the second order contribution to the reduced unitary evolution that
is given by
\begin{eqnarray}
    U^{(2)}_{R\chi}(t) &=& \langle\chi|U^{(2)}(t)|\chi\rangle = \\ \nonumber
	&-&e^{i\frac{N^2g^2}{\omega^2}(\omega t -\sin(\omega t))}
	e^{-i\omega a^\dagger at}e^{\beta(t)a^\dagger-\beta(t)^*a}\times \\ \nonumber
	& &\frac{N\Delta^2}{4}
	\int_0^tdt'e^{-4i(N-1)\frac{g^2}{\omega^2}(\omega t'-\sin(\omega t'))}
	e^{-\alpha(t')a^\dagger+\alpha(t')^*a}\times \\ \nonumber
	& &\int_0^{t'}dt''e^{4i(N-1)\frac{g^2}{\omega^2}(\omega t''-\sin(\omega t''))}
	e^{\alpha(t'')a^\dagger-\alpha(t'')^*a}
\end{eqnarray}
and we have two wildly oscillating integrals. Again we can estimate this term, in
the thermodynamic limit, as being $O\left(\frac{1}{N}\right)$ and then, it is
even much faster in going to zero with respect to the first order term. This proves
our assertion that, in the thermodynamic limit, the unitary evolution is able to
wash out quantum superposition states, if the initial state of the spin bath
is properly set. Besides, we point out that also the second order term of the Dyson
series changes the state of the spin bath due to the product of $H_0$ terms. We do
not report this modified state here as not essential for our discussion.

The main conclusion of this section is that, in the thermodynamic limit,
the leading order unitary evolution tends to an exact solution of the Schr\"odinger
equation. This solution, in the same limit, 
can produce decoherence washing out quantum superposition states.

\section{Discussion and Conclusions}
\label{sec6}

We have showed that a single radiation mode, interacting with an ensemble of two-level
systems, can undergo decoherence during its unitary evolution. The proof is obtained
using perturbation theory and assuming strong coupling between the field and the
ensemble of two-level systems. At the leading order, the unitary evolution
washes out superposition states when the thermodynamic limit is taken and the ensemble
of two-level systems is properly initially prepared. The proof is given both for a
macroscopic quantum superposition state and for a superposition of Fock states.
Higher order corrections till second order are then computed, proving that the
leading order solution tends to be an exact one in the thermodynamic limit. So,
the classical radiation field is an attracting solution when 
a radiation mode interacts with a very large ensemble of two-level atoms.
It is
interesting to note that this solution
of our model seems to be a genuine example of dynamical
collapse of the wave function.

This result opens up the possibility to prove a quite general result. That is,
when $N$ non interacting particles can realize classical evolution of
observables, one may ask if, interacting with such a system and assuming
the proper initial state, such an effect of decoherence may be an ubiquitous one.  
Indeed, our approach proves to be a useful extension of the formal analogy
between statistical physics and quantum mechanics.

\section*{Acknowledgement}

I have to thank Kazuyuki Fujii for many comments that showed to be useful to
improve the paper.

\label{end}

\end{document}